\documentclass[conference]{IEEEtran}

\usepackage[hyphens]{url}
\usepackage{hyperref}
\usepackage{inputenc}
\usepackage[nolist,nohyperlinks]{acronym}
  \acrodef{TC}[TC]{Trusted Computing}
  \acrodef{TCG}[TCG]{Trusted Computing Group}
  \acrodef{TPM}[TPM]{Trusted Platform Module}
  \acrodef{PQC}[PQC]{Post-Quantum Cryptography}
  \acrodef{PQ}[PQ]{Post-Quantum}
  \acrodef{Hash-Based}[HB]{HB}
  \acrodef{RoT}[RoT]{Root of Trust}
  \acrodef{QC}[QC]{Quantum Computer}
  \acrodef{TCB}[TCB]{Trusted Computing Base}
  \acrodef{CRQC}[CRQC]{Cryptographically Relevant Quantum Computer}
  \acrodef{IoT}[IoT]{Internet of Things}
  \acrodef{IMA}[IMA]{Integrity Measurement Architecture}
  \acrodef{ML-KEM}[ML-KEM]{Module-Lattice-Based Key-Encapsulation Mechanism}
  \acrodef{ML-DSA}[ML-DSA]{Module-Lattice-Based Digital Signature}
  \acrodef{SLH-DSA}[SLH-DSA]{Stateless Hash-Based Digital Signature}
  \acrodef{FN-DSA}[FN-DSA]{ FFT over NTRU-Lattice-Based Digital Signature Algorithm}
  \acrodef{HB}[HB]{Hash-Based}
  \acrodef{KEM}[KEM]{Key Encapsulation Mechanism}
  \acrodef{FPU}[FPU]{Floating Point Unit}
  \acrodef{PKI}[PKI]{Public Key Infrastructure}
  \acrodef{CA}[CA]{Certification Authority}
  \acrodef{TEE}[TEE]{Trusted Execution Environment}
  \acrodef{CoT}[CoT]{Chain of Trust}
  \acrodef{TLS}[TLS]{Transport Layer Security}
  \acrodef{TA}[TA]{Trusted Application}
  \acrodef{PCR}[PCR]{Platform Configuration Register}
  \acrodef{fTPM}[fTPM]{Firmware TPM}
  \acrodef{AK}[AK]{Attestation Key}
  \acrodef{EK}[EK]{Endorsement Key}
  \acrodef{REE}[REE]{Rich Execution Environment}
  \acrodef{RTM}[RTM]{Root of Trust for Measurement}
  \acrodef{CRTM}[CRTM]{Core Root of Trust for Measurement}
  \acrodef{RTS}[RTS]{Root of Trust for Storage}
  \acrodef{RTR}[RTR]{Root of Trust for Reporting}
  \acrodef{IETF}[IETF]{Internet Engineering Task Force}
  \acrodef{ROM}[ROM]{Read-Only Memory}
  \acrodef{RA}[RA]{Remote Attestation}
  \acrodef{CoM}[CoM]{Chain Of Measurements}

\usepackage{multirow}
\usepackage{caption}
\usepackage{doi}
\usepackage{siunitx} 
\usepackage{float}
\usepackage{comment}
\usepackage{booktabs}
\usepackage{longtable}
\usepackage{tabularx}
\usepackage{graphicx}
\usepackage{adjustbox}
\begin{document}
\title{Towards Quantum-Resistant Trusted Computing: \\ Architectures for Post-Quantum Integrity Verification Techniques}

\author{\IEEEauthorblockN{Grazia D'Onghia}
\IEEEauthorblockA{Politecnico di Torino\\
Dip. di Automatica e Informatica\\
Torino, Italy\\
grazia.donghia@polito.it}
\and
\IEEEauthorblockN{Antonio Lioy}
\IEEEauthorblockA{Politecnico di Torino\\
Dip. di Automatica e Informatica\\
Torino, Italy\\
antonio.lioy@polito.it}
}
\maketitle
\thispagestyle{plain}
\pagestyle{plain}
\begin{abstract}
Trust is the core building block of secure systems, and it is enforced through methods to ensure that a specific system is properly configured and works as expected.
In this context, a \ac{RoT} establishes a trusted environment, where both data and code are authenticated via a digital signature based on asymmetric cryptography, which is vulnerable to the threat posed by \acp{QC}.
Firmware, being the first layer of trusted software, faces unique risks due to its longevity and difficult update. 
The transition of firmware protection to \ac{PQC} is urgent, since it reduces the risk derived from exposing all computing and network devices to quantum-based attacks.
This paper offers an analysis of the most common trust techniques and their roadmap towards a \ac{PQ} world, by investigating the current status of \ac{PQC} and the challenges posed by such algorithms in existing \ac{TC} solutions from an integration perspective.
Furthermore, this paper proposes an architecture for \ac{TC} techniques enhanced with \ac{PQC}, addressing the imperative for immediate adoption of quantum-resistant algorithms.
\end{abstract}

\IEEEpeerreviewmaketitle

\section{Introduction} \label{sec:introduction}

In digital systems, the concept of \emph{trust} is related to 
\ac{TCB} and \ac{RoT}.
A \ac{TCB} is the totality of protection mechanisms within a computer system (including hardware, firmware, and software), and their combination  is responsible for enforcing a security policy \cite{Tasker2001TrustedCS}, while a \ac{RoT} is a component that performs one or more security-specific functions and is always assumed to behave in the expected manner \cite{tcg_glossary}.
\ac{TC} first became a reality with the \ac{TPM}, which now can be found in millions of laptops and electronic devices around the world. 
Thanks to this component, the building blocks that allow one component in a computer network to trust all other hardware and software related pieces can be established \cite{tc_building_blocks}.
However, since \ac{TC} is based on authentication, that is, digital signatures using public key cryptography, it is threatened by the advent of \acp{CRQC}, as demonstrated by Shor's algorithm \cite{Shor_1997}, capable of factoring large prime numbers in polynomial time.
The urgency of migrating to \ac{PQC} is enhanced by the ``Store now, decrypt later'' attack.
The latter exploits the asymmetry between the time it takes to develop a \ac{CRQC} and the time required to migrate existing systems to \ac{PQ} solutions.
This dynamic is described in Mosca's inequality \cite{mosca}, which warns that systems lacking quantum resistance today risk catastrophic compromise tomorrow.
This threat is particularly acute for firmware protection, where digital signatures must ensure integrity over long device lifecycles.
The US National Security Agency (NSA) underscores this urgency in its Commercial National Security Algorithm (CNSA) 2.0 Suite \cite{cnsa_algo_suite}, explicitly prioritizing post-quantum firmware signing as a critical defence against future quantum adversaries.
Once compromised, firmware vulnerabilities may undermine entire ecosystems, from \ac{IoT} sensors to cloud infrastructure.
This paper directly addresses this imperative, offering actionable recommendations for integrating \ac{PQ} algorithms into firmware signing and attestation workflows, aligning with CNSA 2.0’s call for immediate transitions.
The first essential step in this transition exercise is to evaluate the already standardized \ac{PQC} algorithms proposed by the NIST \cite{nist_standard}.
The purpose of this evaluation is to understand the most suitable algorithm for the specific use case of \ac{TC} and Integrity Verification.
Since \ac{PQC} relies on different mathematical problems and uses keys much larger than current signatures, compatibility must be evaluated with current instruction sets and network protocols.
The contribution of this paper consists in combining the previous recommendations with our requirements to evaluate the best approach towards \ac{PQ} Integrity Verification techniques.
The effort in the \ac{PQ} transition relies in evaluating the overhead introduced by \ac{PQ} algorithms 
and choosing the best ones for two use cases within a Remote Attestation framework.
This paper is structured as follows: section~\ref{sec:tc} contains an overview of the main Integrity Verification techniques,
section~\ref{sec:pqc} summarizes the status of \ac{PQC} standardization with a comparison among standardized algorithms, and section~\ref{sec:challenge} contributes to the \ac{PQ} transition by providing recommendations and requirements on the integration of \ac{PQC} within Integrity Verification techniques.
Finally, Section~\ref{sec:arch} contains our proposed architecture for a system that integrates \ac{PQC} in Integrity Verification, and section~\ref{conclusions} contains the conclusion and examines future works. 
\section{Trusted Computing techniques for Firmware and Software protection}
\label{sec:tc}
Endpoint computers encompass hardware, firmware, drivers, operating systems, and application software, all of which impact the integrity and security of both the devices themselves and the network they are part of.
Modern software and firmware protection rely on three core techniques: secure boot, measured boot, and Remote Attestation.
Such techniques have become an essential requirement, especially to protect \ac{IoT} devices, since attacks against them have become widespread.
The integrity of an \ac{IoT} device system includes load-time integrity (secure and measured boot) and runtime integrity (Remote Attestation) \cite{secure_boot_arm}.
Consequently, there is a difference between integrity verification enforced on the platform (such as secure boot) and Remote Attestation, which allows external entities to verify the integrity status of a device.
Integrity Verification enforced on the platform is realized with the \ac{RTM}, founded in the \ac{CRTM}, which is an immutable portion of the platform initialization code responsible to verify the authenticity of the next entity before transferring control to it.
Meanwhile, Remote Attestation requires the platform to provide three \acp{RoT}: \ac{RTM}, \ac{RTS}, and \ac{RTR}.
\subsection{Secure Boot}
Secure boot ensures that a device starts only with trusted software. 
This is achieved by verifying the digital signature created by the manufacturer on the boot components. 
In this way, a \ac{CoT} is established to ensure the integrity of the load time of each boot partition.
The \ac{CoT} is based on the concept of verifying the next boot image: the image of the former boot partition verifies the image of the next boot partition, and so on until all the boot partitions are successfully loaded. 
The essential aspect of secure boot is that even if a single signature verification fails, the whole  boot process is halted.
Transitioning to a quantum-resistant secure boot involves supporting \ac{PQC} in the boot sequence at the earliest possible level, to protect the whole boot chain from quantum attacks.
Secure boot also requires a careful evaluation of resource requirements and performance, as the \ac{PQ} signature algorithms have to be efficient in the verification operation to keep boot time and system latency low.
\subsection{Measured Boot}
\begin{figure}
    \centering
    \includegraphics[width=0.8\linewidth]{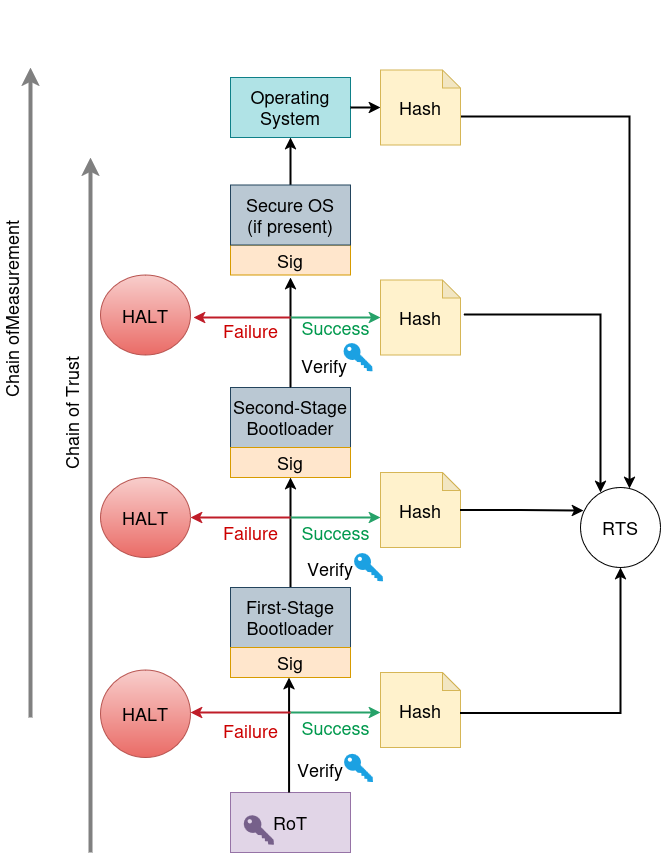}
    \caption{General view of Secure boot and Measured boot.}
    \label{fig:secure_measured}
\end{figure}

Measured boot measures the elements in the \ac{CoT}, from power on until the operating system is fully loaded.
During the boot flow, each critical system component (e.g., firmware, bootloaders, kernel) is measured by the previous one before it gains control of the platform.
The measurement is typically a cryptographic digest computed with a secure hash function, such as SHA-256.
The measured boot chain of trust starts with a \ac{CRTM}, which is typically implemented with a secure and immutable component, contained in \ac{ROM}.
The state of a system changes as programs run with particular configurations.
Measured boot accumulates a list of measurements for each program executed, but does not perform any enforcement like secure boot, which instead halts the system if any attempt is made to execute a program that is not on an approved list \cite{bootstrap_measured_boot}. 
However, both systems involve measuring programs before executing them.
The measurements acquired during the boot stages must be recorded in a secure environment, which usually is the \ac{TPM}, either physical or firmware running inside a \ac{TEE}.
Afterwards, the recorded measurements will be available for use in the Remote Attestation protocol.
Figure~\ref{fig:secure_measured} represents the processes of creating a \ac{CoT} (during secure boot) and a chain of measurement (during the measured boot).
The main difference between them is that secure boot blocks the execution of the boot process by launching a \textit{halt} if a verification fails, while measured boot records the cryptographic hashes of code portions (the measurements), and stores them in the \ac{RTS}. In this way, secure boot creates a chain of trust, while measured boot creates a chain of measurements.

\subsection{Remote Attestation}
Remote Attestation is a protocol used to verify the integrity and trustworthiness of a remote computing system. 
Remote Attestation verifies the integrity and trustworthiness of a remote system (the Attester), which provides reliable information about itself (called evidence) to enable a remote peer (the Relying Party) to decide whether or not to consider that Attester a trustworthy peer.
Additionally, another component called Verifier appraises the evidence via appraisal policies and creates the attestation result to support Relying Parties in their decision process \cite{rats}.
In some architectures, the Relying Party and the Verifier may be merged in a single actor.
Remote Attestation is a challenge-response protocol in which the Verifier sends a challenge (a nonce) to the Attester.
The nonce is used to avoid replay attacks, in which the attacker would send back to the Verifier an old attestation evidence, created before the system is corrupted.
When the attesting system receives the attestation challenge from the Verifier, it generates an attestation evidence, called Quote, which serves as proof that the device's current state (including its firmware, software, and configuration) is trustworthy and has not been tampered with. 
The key elements of a Quote are the measurements performed on the system components, the nonce received from the Verifier and the signature made on the measurement data and the nonce to ensure integrity and authenticity.
In order to enable runtime attestation of a computational node, the \ac{IMA} module, provided by the Linux kernel, must be appropriately configured with a policy.
Finally, 
Attester and Verifier must communicate through secure protocols, such as \ac{TLS} in order to protect the privacy of integrity measurements in transit on the network.
\section{Post-Quantum Cryptography}
\label{sec:pqc}
As of 2024, NIST has finalized four algorithms for standardization: \ac{ML-KEM} \cite{mlkem_standard}, \ac{ML-DSA} (was CRYSTALS-Dilithium), \ac{SLH-DSA} (was SPHINCS+), and \ac{FN-DSA} (was FALCON), whose standard has not been published yet.
Due to the significant differences in implementation, related mathematical problems, memory consumption, and performance, it is essential to differentiate standardized algorithms in order to find the most suitable ones for specific use cases.
 Unlike \acp{KEM}, digital signatures require long-term security guarantees because signatures can be harvested and attacked retroactively once large-scale quantum computers exist \cite{Banerjee}.
This procedure represents the first engineering step for the adoption of \ac{PQC} in existing cryptosystems, and it is as relevant as the mathematical study and implementation of the algorithms themselves.
Standardization is just the first step of actual deployment in existing cryptosystems, and the biggest challenge is including \ac{PQC} in all communication layers and all devices, from clients to service back-ends \cite{inbook}.
 Table~\ref{tab:pqc_comparison} contains a summary of the standardized \ac{PQ} digital signature algorithms with a focus on practical and integration aspects.
 Lattice-Based algorithms leverage complex mathematical problems such as Module Learning With Errors (used in \ac{ML-DSA}) and achieve compact key sizes and an overall efficient verification.
\ac{ML-DSA} supports NIST security levels 2 (\ac{ML-DSA}-44), 3 (\ac{ML-DSA}-65), and 5 (\ac{ML-DSA}-87), offers a balance between speed and security. 
With public keys in the range 1.3-2.6 kB and signatures 2.4-4.6 kB, it is well-suited for every application, but it is recommended for \ac{PKI} and Internet protocols. 
Moreover, Duarte de Menese et al. \cite{mldsa_opt} recently provided an optimization to reduce its computational overhead by shrinking its memory footprint.
However, \ac{ML-DSA} is susceptible to side-channel attacks, thus requiring careful implementation hardening. 
\ac{FN-DSA}, implemented with NIST levels 1 (\ac{FN-DSA}-512) and 5 (\ac{FN-DSA}-1024), distinguishes itself with exceptionally small signatures (0.6–1.3 kB), making it ideal for resource-constrained environments such as \ac{IoT} devices.
Nevertheless, its reliance on 53-bit \ac{FPU} \cite{sok} limits adoption in low-end embedded systems lacking hardware \ac{FPU}.

\ac{HB} signatures are uniquely suited for integrity verification due to their reliance on quantum-secure hash functions rather than structured mathematical problems.
These algorithms prioritize long-term security through collision-resistant hash functions, though they introduce a significant overhead in performance.
\ac{HB} schemes are recommended mostly for safeguarding integrity-sensitive systems, particularly in domains that require high security such as firmware signing, secure boot \cite{cnsa_algo_suite}, and \acp{CA} \cite{shaping_qr_future_pki}. 
Among these, \ac{SLH-DSA} stands out for its stateless design, which does not involve state management.
The stateless nature of SLH-DSA simplifies deployment in decentralized architectures, but its operational costs limit adoption to niche, high-security domains. 
\ac{SLH-DSA}’s large signature sizes (7.8–49.8 kB) and slow signature generation make its adoption for real-time applications a big challenge.
In contrast, stateful \ac{HB} schemes like XMSS and LMS produce smaller signatures (2.5–12 kB) and adhere to RFC standards \cite{rfc8391}.
These algorithms are optimized for use cases such as secure boot chains and firmware signing.
LMS, in particular, demonstrates strong performance on embedded systems \cite{Campos2020LMSvsXMSS}, but both XMSS and LMS require robust state management to mitigate key-reuse risks, thus requiring solutions such as \acp{TEE} to manage the state.
Despite these operational complexities, stateful \ac{HB} schemes are interesting options for the \ac{PQ} transition of \ac{TC} architectures, especially in resource-constrained devices.

Both lattice-based and \ac{HB} algorithms can be implemented in \ac{TC}, but \ac{HB} algorithms rely on the well-established security of the underlying hash functions, thus being more suitable for such a high security purpose.
 Despite providing the same NIST security levels, it is worth noting that the security of lattice-based algorithms has been investigated less since they are based on novel mathematical problems.
\begin{table*}[htb]
\caption{Qualitative overview of Lattice-Based and Hash-Based Post-Quantum Algorithms}
\label{tab:pqc_comparison}
\Large
\begin{adjustbox}{width=\textwidth}
\begin{tabular}{|l|l|l|l|l|l|}
\hline
\textbf{Algorithm} & \textbf{NIST security Levels} & \textbf{Key/Signature Sizes} & \textbf{Strengths} & \textbf{Weaknesses} & \textbf{Best Use Case} \\
\hline
\multicolumn{6}{|l|}{\textbf{\textit{Lattice-Based}}} \\
\hline
\ac{ML-DSA} & 2, 3, 5 & 
\begin{tabular}[t]{@{}l@{}}
Pub: 1,312--2,592 B \\ 
Priv: 2,560--4,896 B \\ 
Sig: 2,420--4,627 B
\end{tabular} 
& 
\begin{tabular}[t]{@{}l@{}}
- Balance of speed and security \cite{Banerjee} \\ 
- Optimizations made \cite{mldsa_opt} 
\end{tabular} & 
\begin{tabular}[t]{@{}l@{}}
- Large signatures \\ 
- Vulnerable to \\
side-channel attacks \cite{Bernstein2017}
\end{tabular} & 
\begin{tabular}[t]{@{}l@{}}
- \ac{PKI} \\
- Internet applications
\end{tabular} \\
\hline
\ac{FN-DSA} & 1, 5 & 
\begin{tabular}[t]{@{}l@{}}
Pub: 897--1793 B \\ 
Priv: 1,281--2,305 B \\ 
Sig: 666--1,280 B
\end{tabular} & 
\begin{tabular}[t]{@{}l@{}}
- Smallest signatures \\ 
- Efficient verification
\end{tabular} & 
\begin{tabular}[t]{@{}l@{}}
- Requires 53-bit \ac{FPU}
\end{tabular} & 
\begin{tabular}[t]{@{}l@{}}
- Resource-constrained \\
environments \\
- IoT devices
\end{tabular} \\
\hline
\multicolumn{6}{|l|}{\textbf{\textit{Hash-Based}}} \\
\hline
\ac{SLH-DSA} & 1, 3, 5 & 
\begin{tabular}[t]{@{}l@{}}
Pub: 32--64 B \\ 
Priv: 64--128 B \\ 
Sig: 7,856--49,856 B
\end{tabular} & 
\begin{tabular}[t]{@{}l@{}}
- Stateless \\ 
- Conservative security
\end{tabular} & 
\begin{tabular}[t]{@{}l@{}}
- Massive signatures \\ 
- Slow signing/verification
\end{tabular} & 
\begin{tabular}[t]{@{}l@{}}
- Long-term integrity \\ 
(e.g., firmware, logs) \\
- High security domains \\
(e.g., \ac{CA}) \cite{shaping_qr_future_pki}
\end{tabular} \\
\hline
XMSS & 1, 3, 5 & 
\begin{tabular}[t]{@{}l@{}}
Pub: 32--64 B \\ 
Priv: 1,088--2,560 B \\ 
Sig: 2,500--12,000 B
\end{tabular} & 
\begin{tabular}[t]{@{}l@{}}
- Stateful but smaller \\ 
signatures than \ac{SLH-DSA} \\ 
- RFC 8391 standardized \cite{rfc8391}
\end{tabular} & 
\begin{tabular}[t]{@{}l@{}}
Requires secure \\ 
state management
\end{tabular} & 
\begin{tabular}[t]{@{}l@{}}
- Secure boot \ac{CoT} \cite{pq_secureboot}\\
- Firmware signing \cite{cnsa_algo_suite}
\end{tabular} \\
\hline
LMS & 1, 3, 5 & 
\begin{tabular}[t]{@{}l@{}}
Pub: 32--64 B \\ 
Priv: 1,088--2,560 B \\ 
Sig: 2,500--12,000 B
\end{tabular} & 
\begin{tabular}[t]{@{}l@{}}
- Stateful but smaller \\ 
signatures than \ac{SLH-DSA} \\ 
- Best performance among \ac{HB} \\
on embedded systems \cite{Campos2020LMSvsXMSS}
\end{tabular} & 
\begin{tabular}[t]{@{}l@{}}
- Requires secure \\ 
state management
\end{tabular} & 
\begin{tabular}[t]{@{}l@{}}
Secure boot \ac{CoT} \cite{pq_secureboot_hb}\\ 
\end{tabular} \\
\bottomrule
\end{tabular}
 \end{adjustbox}
\end{table*}

\section{Challenges of \ac{PQC} integration in integrity verification techniques} \label{sec:challenge}
Migrating Integrity Verification techniques to \ac{PQC} is a complex, yet essential transition exercise that demands a thorough revision of key management practices, cryptographic libraries, and protocol standards used into existing frameworks.
This section analyzes system requirements and designs building blocks needed to carry out transition of Integrity Verification techniques to \ac{PQC} on different device categories: embedded systems and network platforms.

\subsection{Secure Boot}
Making secure boot quantum-resistant requires first \ac{PQC} integration at the firmware level, to ensure that the initial code executed during the boot process is authenticated using a \ac{PQ} signature. 
Some key factors must be considered when choosing the optimal digital signature algorithm to use for \ac{PQ} secure boot. 
First, \ac{PQ} digital signatures used for secure boot require a high security strength and algorithm maturity, as they should remain valid for a long period of time.
The security of \ac{HB} schemes relies solely on the properties of the underlying hash functions, which are well-studied and understood in cryptography, providing a solid foundation for their security compared to newer approaches, such as lattice-based ones \cite{Bernstein2015}.
This consideration leads to preferring \ac{HB} schemes in the implementation of \ac{PQ} secure boot.
Among \ac{HB} schemes, NIST selected \ac{SLH-DSA} for standardization, while the \acs{IETF} developed RFCs for two Stateful \ac{HB} schemes, LMS \cite{rfc8554} and XMSS \cite{rfc8391}. 
Furthermore, it is crucial to ensure that the integration of \ac{PQ} algorithms does not introduce new vulnerabilities in the secure boot process.
Stateful \ac{HB} algorithms (LMS, XMSS) require secure and reliable mechanisms for state storage and updates to ensure their security and correct operation. Therefore, their adoption would add complexity.
However, when these algorithms are used to implement secure boot, they do not increase the implementation complexity in the device, as the management and updating of the state have to be carried out by the entity that signs the software, which is external to the device.
Another critical aspect of the secure boot process is the management of cryptographic keys.
\ac{PQ} key management involves first generating \ac{PQ} public and private keys by the firmware and software provider. 
The private key is used to sign the firmware and software, so it has to be stored securely by the software provider and never shared. 
The public key is instead used during the device boot to verify the authenticity and integrity of the boot components, so it has to be distributed and embedded in the device's firmware, or in its \ac{RoT}.
If a private key is compromised, a revocation mechanism must be in place to invalidate the old key and prevent its use.
This often involves updating the firmware with a new public key.
Therefore, \ac{PQ} secure boot should support a secure firmware update mechanism that includes updating the embedded public key.
When choosing the \ac{PQ} signature algorithm, it is necessary to take into account that the adoption of stateful \ac{HB} schemes would add complexity to private key management, because the signer must keep track of the used and remaining private key components. 
This state must be securely stored and updated after each signature generation to prevent reuse of the same components, which would compromise security.
There are several works that demonstrate the feasibility of \ac{PQ} secure boot through the adoption of \ac{HB} algorithms.\cite{secureboot_pq_era} presents a comprehensive evaluation of signature verification of the standardized NIST \ac{PQ} algorithms, showing that \ac{HB} algorithms have the best performance.
XMSS algorithm is adopted for secure boot in \cite{pq_secureboot}, offering great performance and reduced signature sizes. Other \ac{HB} schemes such as LMS and SPHINCS+ are evaluated in \cite{pq_secureboot_hb}.
\subsection{Measured Boot}
Implementing a \ac{PQ} measured boot requires adopting quantum-resistant hash functions to generate cryptographic hashes of code and configurations.
The impact of a \ac{CRQC} on hash functions is less severe than the impact on digital signatures, but still significant.
The security strength of a hash function against quantum attacks is primarily evaluated based on Grover's algorithm \cite{grover1996fast}, which can speed up the brute-force search for pre-image attacks, effectively reducing the security level of the hash function by half \cite{Mina2020ThreatsTM}. 
The best known quantum attack does not improve significantly over the classical birthday attack, which halves the security strength \cite{Bernstein2017}.
Therefore, to implement a \ac{PQ} measured boot, hash functions like SHA-384, SHA-512, or SHA-3-512 should be used, ensuring sufficient collision resistance against quantum threats.
Then, the recorded quantum-resistant measurements are available to be used in the \ac{PQ} remote attestation protocol. 
\subsection{Remote Attestation}
Remote attestation requires some considerations about \ac{PQC} integration, specifically for the hash and  signatures algorithms employed.
To enable measurement of files accessed at system runtime with quantum-resistant hash algorithms, the \ac{IMA} module has to be configured to use these algorithms (e.g. SHA-512, available since kernel version 3.13, or SHA-3-512 available since version 6.7).
Furthermore, to protect the integrity of the IMA Log file with SHA-512 or SHA-3-512, a \ac{TPM} must be present in the system and it must be configured with a \ac{PCR} bank associated with SHA-512 or SHA-3-512.
In the case of \ac{PQ} remote attestation, the selected signature algorithm should be efficient in all its operations, especially for signature generation and verification. 
The efficiency of these operations is crucial for quickly detecting compromises and maintaining overall system performance: the faster the attestation process, the shorter the window of opportunity for attackers to exploit a compromised system.
An adoption of \ac{PQ} algorithms in Remote Attestation can be found in \cite{demonstrating_pq_ra}, where Dilithium2 and FALCON are evaluated against ECDSA.
In this work, FALCON represents an optimal choice for \ac{PQ} remote attestation, because it provides the highest efficiency in signature generation and verification. 
Since different evaluations can be made for remote attestation based on the use case, enabling algorithm agility for the attestation evidence signature is a desirable feature.
\section{Proposed Architecture} \label{sec:arch}
\begin{figure}
    \centering
    \includegraphics[width=0.9\linewidth]{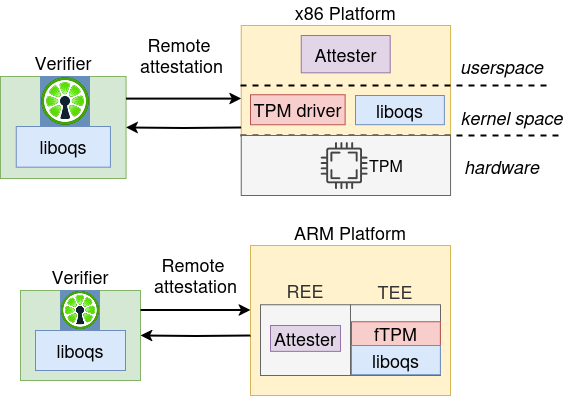}
    \caption{High level architecture of the proposed solution.}
    \label{fig:arch}
\end{figure}

The architecture proposed here finalizes the investigation that has been done about \ac{PQC} and its integration into firmware and software protection techniques.
However, this is the first step in the complete transition of cryptosystems, which includes the full hardware integration of \ac{PQ} standardized algorithms.
At its core, the architecture operates on two parallel fronts, each tailored to a distinct deployment environment. 
Figure~\ref{fig:arch} represents the high level architecture of the framework.

For modern ARM-based embedded systems, trust is anchored in a \ac{PQ} \ac{fTPM} running within ARM TrustZone. 
This \ac{fTPM} includes \ac{PQ} 
signatures to sign the attestation Quote 
, ensuring quantum-resistant integrity verification.
This solution addresses an use case where the \ac{PQ} transition must be done on a system without a physical \ac{TPM}, thus using a firmware \ac{TPM}.

On the other side, for x86 systems reliant on physical \ac{TPM}, the architecture adopts a hybrid model that combines classical ECDSA signatures with PQ wrappers at the kernel level. 
After receiving the attestation request from the Verifier, the Attester asks the physical \ac{TPM} to generate the Quote, which is first signed with ECDSA and then wrapped with a \ac{PQ} signature by an extension in the \ac{TPM} driver, obtaining a hybrid Quote, 
thus adapting to this first phase of the transition.
Meanwhile, the Verifier enhances the Keylime framework \cite{keylime} with \ac{PQC} support via liboqs library \cite{liboqs}, in order to understand the \ac{PQ} Quote received from the Attester, whether it is ARM or x86, and to be able to perform signature verification.

The key difference in the two flavours is the generation of the Quote.

For the ARM platform it is important to highlight the difference between the \ac{REE} (namely, the Normal World in ARM TrustZone) and the TEE.

On the other side, the meaningful separation that is drawn in the x86 platform is between userspace (where regular applications run), kernel space (with kernel drivers that interact directly with the hardware), and the hardware itself, where the physical \ac{TPM} is located.
Acknowledging the phased nature of \ac{PQ} adoption, the architecture prioritizes interoperability.
Hybrid workflows allow legacy systems to coexist with \ac{PQ}-enhanced  components.

On ARM platforms, the \ac{fTPM} demonstrates how constrained devices can adopt \ac{PQ} primitives in integrity verification without needing a physical \ac{TPM}. 
On x86 systems, kernel-level \ac{PQ} wrappers provide an assurance for environments where hardware upgrades are impractical. 
Together, these strategies close critical gaps in today’s trust chains while preparing the ground for fully \ac{PQ}-native ecosystems.
This solution is only at design stage, and the evaluation of the implementation and performance of the adopted algorithms will be done in another research activity.
However, when dealing with runtime operations such as Remote Attestation, security strength must be paired with performance, especially in signature generation (involved in the creation of the Quote), therefore the optimal algorithm for our solution is \ac{ML-DSA}.
\section{Conclusion and Future Work} \label{conclusions}
Quantum computers will force a paradigm shift in how we safeguard digital system, especially  firmware and software ecosystems which need long-term security. 
By proposing a layered architecture that integrates \ac{PQC} across the different steps of integrity verification, this paper addresses a critical gap in today’s cybersecurity landscape: the lack of end-to-end quantum-resistant trust chains.
Current architectures often treat secure boot, measured boot, and Remote Attestation as isolated processes, creating fragmented trust boundaries that could be eventually exploited by quantum adversaries.
This work redefines integrity verification as a cohesive, layered framework where \ac{PQC} is included into every stage of trust propagation, thus making a foundational step toward cryptographic continuity, where systems adapt dynamically to emerging threats.
This  design bridges the divide between legacy systems and emerging \ac{PQ} standards, offering a dual-path strategy, for modern ARM-based platforms and legacy x86 systems, that balances innovation with pragmatic transition needs.
While this paper outlines an important draft, the journey to practical, scalable \ac{PQ} adoption has only begun.
Future works include rigorous testing and evaluation of the proposed architectures to quantify their real-world effect.
Future research should explore the development of lightweight \ac{PQ} primitives optimized for firmware protection, such as \ac{HB} signatures with smaller footprints or lattice-based schemes with reduced computational overhead.
As quantum computing transitions from theory to reality, the lessons of this paper are clear: firmware protection is a building block of global cybersecurity, consequently its \ac{PQ} transition is inevitable.
The designs proposed here are not endpoints but basis for a collective, urgent effort to quantum-resistant trust architectures.
Implementation details, empirical validations, and policy frameworks will follow in subsequent work, but the concept is very clear: the time to act is now.

{\footnotesize
\textbf{Acknowledgments.}
This work is part of the QUBIP European project -- \url{https://qubip.eu/} -- funded by the European Union under the Horizon Europe framework programme (grant agreement no. 101119746).
}
\bibliographystyle{torsec}
\bibliography{biblio}

\end{document}